# All in action

ARTO ANNILA[1,2,3,*]

[1]*Department of Physics*, [2]*Institute of Biotechnology* and [3]*Department of Biosciences, FI-00014 University of Helsinki, Finland*

The principle of least action provides a holistic worldview in which nature in its entirety and every detail is pictured in terms of actions. Each and every action is ultimately composed of one or multiples of the most elementary action which corresponds to the Planck's constant. Elements of space are closed actions, known as fermions, whereas elements of time are open actions, known as bosons. The actions span energy landscape, the Universe which evolves irreversibly according to the 2$^{nd}$ law of thermodynamics by diminishing density differences in least time. During the step-by-step evolution densely-curled actions unfold by opening up and expelling one or multiple elementary actions to their surrounding sparser space. The manifold's varieties process from one symmetry group to another until the equivalence to their dual, i.e., the surrounding density has been attained. The scale-free physical portrayal of nature does not recognize any fundamental difference between fundamental particles and fundamental forces. Instead a plethora of particles and a diaspora of forces are perceived merely as diverse manifestations of a natural selection for various mechanisms and ways to decrease free energy in the least time.

## 1. Introduction

Actions integrate momenta along paths of an energy landscape (1,2,3). The powerful principle of least action delineates flows of energy on least-time trajectories. These geodesics can be determined when the manifold remains invariant under the influence of action whereas this task turns out to be intractable when the flows drive the energy landscape in evolution. Even so, physics is able with its most general concepts to tackle such problematic processes too.

Nature can be pictured in its entirety and every detail as a landscape that embodies diverse densities of energy. When energy flows from a spatial density to another, the manifold will morph itself so that continuity and conservation of energy are satisfied. Specifically, when currents circulate on closed, bound orbits, the local manifold retains its symmetry over the conserved motional period (4). In general, when currents spiral along open, unbound paths, the landscape evolves irreversibly by breaking its steady-state symmetry for another via non-conserved transformations. The laws of motion can be expressed concisely as actions in classical electromagnetism (5), general relativity (6) and quantum electrodynamics (7) but perhaps it is less appreciated that also the 2$^{nd}$ law of thermodynamics (8) can be formulated accordingly to describe evolutionary processes.

Customarily the 2$^{nd}$ law is written as a differential transformation from a state toward another, more probable one, but the universal law can also be cast in an integral form. Then the least action conveys the system from one state to a more probable one at the maximal rate. In other words, the principle of increasing entropy and the principle of least action are equivalent imperatives (9). The natural law for the maximal energy dispersal accounts for diverse irreversible processes that consume density differences, i.e., free energy in least time (10,11,12). Eventually, when all forms of free energy have been exhausted, the open evolutionary paths will close to the optimum orbits of a conserved stationary state. Then the landscape is even and can be characterized by a gauge symmetry group. Thus, the evolutionary equation derived from the statistical physics of open systems (13,14) offers insights as well to the motions of Hamiltonian systems. The objective of this study is to show how some familiar forms of physics unite when nature is described in a comprehensive and self-similar manner as actions within surrounding actions.

## 2. The natural law of maximal energy dispersal

The 2$^{nd}$ law of thermodynamics simply says that differences in energy densities will level off in least time. Consumption of free energy via the most voluminous flows of energy powers evolution from one state to another. Motion along the emerging optimal path proceeds along the steepest descent in time ($\partial_t = \partial/\partial t$) which is equivalent to the steepest directional (i.e., velocity $v$) gradient ($D = \mathbf{v}\cdot\nabla$). The statistical measure for a natural process is the logarithmic, additive probability known as entropy $S = k_B \ln P$ (15).

According to the natural law entropy will not only be increasing but it will be increasing as quickly as possible. This universal imperative is also known as the maximum entropy production principle (16,17), the maximum power principle (18) and the principle of least curvature (19).

The principle of increasing entropy as the evolutionary equation of motion is obtained from the statistical physics of open systems (13,14)

$$d_t S = k_B\, d_t P/P = k_B L \geq 0, \quad (2.1)$$

where the rate of entropy change is proportional by Boltzmann's factor $k_B$ to the process generator $L = k_B T d_t Q$ in accordance with classical thermodynamics (8). The probability in motion $d_t P = LP$ (Eq. 2.1) is expressed using Gibbs' formalism for energy densities (20). Each energy density $\phi_j$, present in indistinguishable numbers $N_j$ of $j$-entities, is assigned with $\phi_j = N_j \exp(G_j/k_B T)$ where $G_j$ is relative to the average energy $k_B T$ of the system per entity. According to the scale-free formalism (21,22) each $j$-entity itself is a system of diverse $k$-entities. Each population $N_k$, in turn, is associated with $\phi_k = N_k \exp(G_k/k_B T)$. These bound forms of energy are the material entities (fermions) that exclude each other in space. Their interactions as flows of energy are communicated over time by force-carrier quanta (bosons).

The entire system is composed of diverse systems within systems (23) (Fig. 1). It is summarized by the total probability (13,14)

$$P = \prod_{j=1} P_j = \prod_{j=1}\left(\prod_k \left(N_k e^{-(\Delta G_{jk} - i\Delta Q_{jk})/k_B T}\right)^{g_{jk}} \bigg/ g_{jk}!\right)^{N_j} \bigg/ N_j ! \quad (2.2)$$

defined in a recursive manner so that each $j$-entity, in indistinguishable numbers $N_j$, is a product $\Pi N_k$ of embedded $k$-entities, each distinct type available in numbers $N_k$. The energy difference between the $j$ and $k$ entity is $\Delta G_{jk} = G_j - G_k$ or $g_{jk}\Delta G_{jk} = G_j - g_{jk}G_k$ when the $j$-entity forms from indistinguishable (symmetrical) $k$-entities in degenerate numbers $g_{jk}$. The change in boson vector potential, i.e., radiation $\Delta Q_{jk}$ couples to the $jk$-transformation orthogonal, as indicated by $i$, to the fermion scalar potential difference.

When the total logarithmic probability is multiplied by $k_B$

$$\ln P = \sum_j \ln P_j \approx \sum_j N_j\left(1 - \sum_k (\Delta \mu_{jk} - i\Delta Q_{jk})/k_B T\right) \quad (2.3)$$

the aforementioned additive statistical measure $S$ for the entire system is obtained (24). The free energy $A_{jk} = \Delta\mu_{jk} - i\Delta Q_{jk}$, known also as affinity (25), is the motive force that directs the transforming flow $d_t N_j$ from $N_k$ to $N_j$ by the scalar $\Delta\mu_{jk} = \mu_j - \Sigma\mu_k = k_B T(\ln\phi_j - \Sigma\ln\phi_k)$ and vector $\Delta Q_{jk}$ potential differences. The adopted approximation $\ln N_j! \approx N_j \ln N_j - N_j$ implies that $\ln P_j$ is a sufficient statistic (26) for $k_B T$ to characterize the $j$-system so that it may absorb or emit quanta without a marked change in its average energy content, i.e., $A_{jk}/k_B T \ll 1$. The system holds capacity $C = TdS/dT$ in its diverse populations $N_j$ and in the free energy terms $A_{jk}$ to resist changes in its average energy imposed by surroundings at a different temperature. The energy density difference, i.e., force means that the curvature of the system's energy landscape differs from that of its surrounding system. The formula for specific heat $C = k_B \partial \ln P/\partial \ln T$ is also known as the renormalization group equation (27,28,29). Usually $C > 0$ but when $\Delta Q$ exceeds $\Delta\mu$ then the high-energy characteristic $C < 0$ manifests itself as asymptotic freedom.

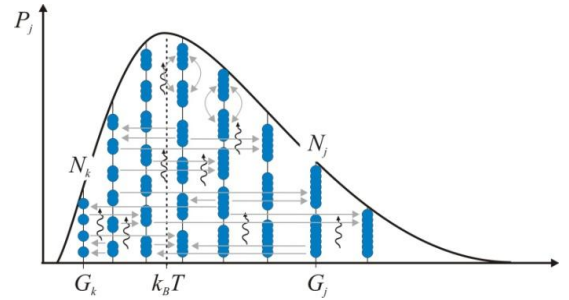

Figure 1. A self-similar energy level diagram describes nested hierarchy of nature where each $j$-system (a composite solid in blue color) is regarded as a system within systems that are all being ultimately composed of multiple elementary constituents (blue single solids at most left). All systems evolve via step-by-step $jk$-transformations toward more probable states by consuming mutual density differences $\Delta\mu_{jk}$ (horizontal arrows) and those $\Delta Q_{jk}$ (vertical wavy arrows) relative to the surroundings in least time. Isergonic exchange (bow arrows) account for reversible processes that do not affect the average energy $k_B T$. At any given time the probability distribution $P_j$ outlines the maximum entropy partition of a sufficiently statistical system.

A system that does not have enough capacity to resist changes, will evolve abruptly. At these critical events (30), e.g., when a new $j$-species emerges or an old one goes extinct, $P_j$ will change at once, e.g., from fifty-fifty indeterminism to full certainty. Moreover, when interactions are insufficient to establish common $k_B T$ over a given period $\tau$ of time, the entities fail to form a system but remain as constituents that surround sufficiently statistical systems at a lower level of hierarchy where interactions are more frequent and intense (31,32,33).

According to the natural law, given as an equation of evolution (Eq. 2.1), a system which is higher in energy density than its surroundings, will evolve from its current state to a more probable one by displacing quanta to the sparse surroundings. Conversely, a system which is lower in energy density than its respective surroundings, will evolve by acquiring quanta from the dense surroundings. Any two states are distinguishable from each other only when the $jk$-transforming flow is dissipative $\Delta Q_{jk} \neq 0$ (14,34). In view of that a net non-dissipative system is stationary. The maximum entropy state is Lyapunov-stable so that an internal perturbation $\delta N_j$ away from the steady-state population $N_j^{ss}$ will induce returning forces and opposing flows, i.e., $dS(\delta N_j) < 0$ and $d_t S(\delta N_j) > 0$ (35). In contrast any change in surroundings will compel the system to move toward a new steady state according to the Le Chatelier's principle (36).

The evolutionary equation of motion (Eq. 2.1) is obtained from Eq. 2.3 by differentiating $(\partial P_j / \partial N_j)(dN_j / dt)$ (13,14)

$$d_t \ln P = L = -\sum_{j,k} d_t N_j \frac{A_{jk}}{k_B T} \geq 0 \qquad (2.4)$$

when the time step $\Delta t$ is denoted as continuous $dt$. The notion of continuous motion is in accordance with entropy being a sufficient statistic for $k_B T$. However, the actual $jk$-transformations do advance in quantized steps of $\Delta Q_{jk}$ during $\Delta t$. The resulting changes in $P$ (Eq. 2.2) due to diminishing density differences between the system and its surroundings correspond to a step-by-step rather than a continuous change toward a common $k_B T$.

## 3. Evolving energy landscape

The equation of motion for the evolving energy landscape is developed by multiplying Eq. 2.4 with $k_B T$ to give the continuity for the flows of energy (14,34)

$$k_B T d_t \ln P = -\sum_{j,k} d_t N_j \Delta \mu_{jk} + i \sum_{j,k} d_t N_j \Delta Q_{jk} \qquad (3.1)$$

which result from changes in the scalar and vector potentials. Using the definitions $\mu_j = \partial_{N_j} U_{jk}$ and $d_t N_j = v_j \partial_x N_j d_t x$ and $v_j = d_t x_j$ the three-term formula is transcribed to a convenient continuum approximation

$$\sum_{j,k} d_t 2K_{jk} = -\sum_{j,k} v_j \partial_{x_j} U_{jk} + i \sum_{j,k} \partial_t Q_{jk} \qquad (3.2)$$

where the changes in kinetic energy $2K$ and scalar $U$ and vector $Q$ potentials are given in the Cartesian base of space $j,k = \{x,y,z\}$ and time $t$. The flow equation simply says that when the irrotational potential energy $U_{jk} = -x_j m_{jk} a_k$ is consumed during $dt$, the power $\partial_t Q_{jk} = v_j \partial_t m_{jk} v_k = v_j \partial_t E_{jk} v_k / c^2$ is dissipated and the balance is maintained by the change in the kinetic energy $2K_{jk} = v_j m_{jk} v_k$ (Fig. 2). This has been conjectured already a long time ago (37,38) and also given by Cartan's magic formula (39). However, customarily the general formalism is constrained to topology-preserving homeomorphic Hamiltonian flows to enforce computability.

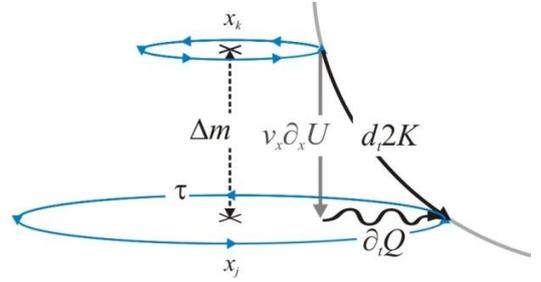

Figure 2. Evolution places a closed action coordinated at $x_k$ to another bound at $x_j$ so that the change in kinetic energy $d_t 2K$ balances the changes in the scalar $v_x \partial_x U$ and vector potentials $\partial_t Q$. The concomitant change in mass $\Delta m = E/c^2$ equals dissipation to the surroundings. At the stationary state net fluxes vanish, so that the closed least-action trajectory $d_t 2K = 0$ can be integrated over the period of motion $\tau$ to the virial theorem $2K + U = 0$.

When the system is in a steady state, the divergence-free part of the force, i.e., the net dissipation vanishes. Then the energy content of the stationary landscape spanned by the $j$- and $k$-entities is denoted by the invariant inertia $x_j m_{jk} x_k$ and by the invariant mass $m_{jk}$. According to $m = E/c^2$ the mass defines the $jk$-system's energy content in terms of a radiation equivalent which is dissipated, i.e., absorbed into the surrounding energy density, the free space known as the vacuum. The systemic energy is in relation to its surrounding radiation by the index $n_{jk} = c^2/v_j v_k$ (isotropic $v_j = v_k$). For example, high masses of gauge bosons $W^\pm$ and $Z^0$ signal that they are themselves sources of dissipation via decay and scatter processes involving leptons. Conversely, energy in radiation when it is spatially confined to a standing wave can be given in terms of a mass equivalent. This conservation of energy is the fundamental equivalence of fermion-to-boson transformations.

The notion of a manifold in motion becomes more vivid when the dissipative flow from the density at $x_k$ toward the density at $x_j$ during $t$ is pictured to funnel along an arc $s_{jk}$. The affine connection along a continuous curve $x = x(t)$ between the two spatial density loci spans a length $s = \int (\mathbf{F} \cdot \mathbf{v})^{1/2} dt = \int (d_t 2K)^{1/2} dt$ where the integrand is referred to as the Rayleigh-Onsager dissipation function (40) or as $Td_tS$ by Gouy and Stodola (41,42). In other words, the manifold's stationary varieties are the integrable conserved currents over the corresponding orbital period $\tau$. In contrast evolutionary elements are non-integrable flows of energy over time $t$ and their dependence on path is appropriately denoted by the inexact differential $đ_t$. A small flow will not perturb much a sufficiently statistical system but it will move a microscopic system substantially, eventually to a state beyond recognition. An energy flow, such as light, from the system to its observing surroundings (or vice versa) underlies detection of any kind and causality in general (14).

The concise notation for the many motions of the differentiable manifold (Eq. 3.2) is developed further by denoting the spatial $\partial_x$ and temporal $\partial_t$ gradients as a 4-vector (43,44)

$$\partial_\mu = (-\partial_t/c, \nabla) = (-\partial_t/c, \partial_x, \partial_y, \partial_z). \qquad (3.3)$$

When $\partial_\mu$ acts on the scalar $U$ and vector $\mathbf{Q}$ potentials that are in turn given as the free energy 4-vector potential in the one-form space-time basis

$$A_\mu = (-U, \mathbf{Q}) = (-U, Q_x, Q_y, Q_z), \qquad (3.4)$$

the curvature in the two-form $\mathbf{F} = d\mathbf{A}$ is obtained. It is represented by the covariant antisymmetric rank 2 tensor

$$F_{\mu\nu} = \partial_\mu A_\nu - \partial_\nu A_\mu = \begin{pmatrix} 0 & -F_x & -F_y & -F_z \\ F_x & 0 & R_z & -R_y \\ F_y & -R_z & 0 & R_x \\ F_z & R_y & -R_x & 0 \end{pmatrix} \qquad (3.5)$$

where the translational and rotational changes in momentum are $d_t\mathbf{p} = \mathbf{F} = -\nabla U + \partial_t\mathbf{Q}/c$ and $\mathbf{R} = \nabla \times \mathbf{Q}$. The form is familiar as the Lorentz force when the components of scalar $\phi = U/q$ and vector $\mathbf{A} = -\mathbf{Q}/qc$ potentials are divided by charge $q$. The change $d_t\mathbf{p}$, i.e., force $\mathbf{F}$ relates to the change in the angular momentum $d_t\mathbf{L}$, i.e., torque $\boldsymbol{\tau} = \mathbf{r} \times \mathbf{F}$ normalized by radius of curvature $\mathbf{r}$ (Fig. 2). The continuity is contained in the invariants $F_{\mu\nu}F^{\mu\nu} = 2(\mathbf{F}^2 - \mathbf{R}^2)$ and $*F_{\mu\nu}F^{\mu\nu} = 4\mathbf{F} \cdot \mathbf{R}$ and $*F_{\mu\nu}*F^{\mu\nu} = 2(\mathbf{F}^2 + \mathbf{R}^2)$. These fixed magnitudes remind us of the three sides of the right angle that is generated from two integers $f$ and $r$ (45,46). Thus the Pythagorean theorem for the three invariants implies that the Euclidean manifold (47,48), formalulated here as continuously differentiable, is in fact modular.

Specifically, at the stationary state the exterior derivate of the system's curvature yields the conserved currents by $d*\mathbf{F} = \mathbf{J}$ so that the conserved energy density, i.e., invariant mass, orbits with phase velocity $d_t\varphi = \omega$ exactly once in a period $\tau = \omega^{-1}$ on the closed least-action path according to $\mathbf{p} \times \boldsymbol{\omega} = -\nabla U$. At the thermodynamic steady state there is no net emission from the system or net absorption to the surroundings over the motional period. Then the exterior algebra of the system's dual $d\mathbf{F} = 0$ says the surrounding is a flat landscape with respect to the system and thus exerts no forces. In other words, the system's average energy density $k_BT$ matches exactly that of the surroundings. At the stationary state there is no net curvature and light propagates straight according to the Bianchi identity. The Maxwell's equation holds for any homogenous medium. The surrounding vacuum has no means to evolve a step further down because the electromagnetic radiation with its symmetry group U(1), the most elementary one cannot be broken down any further. This is to say that the photon has no mass. In quantum electrodynamics the massless gauge boson communicates exchange of energy in interactions that conserve the stationary state. A Hamiltonian system remains steady ($d\ln P = 0$) over the characteristic period of its motions since transformations by the Abelian symmetry group do not introduce energy gradients. These reversible flows are the familiar conserved currents along tractable trajectories that are straight geodesics of even landscapes (4). The closed, parallel actions as Euclidean varieties comply with the ordering relations. Therefore the steady states are countable and any-one closed ring as a subset of affine space is an algebraic variety (49,50).

In general, the evolving energy landscape, represented by the curvature 2-form $\mathbf{F} = d\mathbf{A}$ of some principal high-symmetry bundle, is leveling off via symmetry-breaking transformations from any state to a more probable one. Bound forms of energy, i.e., fermions open up to output free forms of energy, i.e., bosons that dissipate to the surroundings. Since the system and its surroundings share a common interface for the flows of energy, both the system and its surroundings will increase in entropy when mutual energy density differences are decreasing. The conservation of energy including both the system and its dual is respected by the differential geometry where the Hodge star operator, in accordance with metric signature (+,+,+,–), transforms an oriented inner product density in an element of space, e.g., given in the Cartesian base, $*(dx \wedge dy \wedge dz) = dt$ to radiated density carried away by an element of time where the exterior product is a basis-independent formulation of volume.

It is worth noting that transformations of the Lorentz group SO(3,1), although being conveniently continuous isometries of Minkowskian spacetime (51), are conserved but not evolutionary processes. When no symmetry is broken, no quantum is dissipated and no change of state has happened. The sesquilinear inner product defines the stationary-state unitary space with the Euclidean norm which is invariant under multiplication by the complex numbers of a unit norm. In these special cases, the equation of motion describes merely an isergonic phase precession that can be solved by a unitary transformation. In contrast the equation of evolution cannot be solved when a natural process with three or more degrees of freedom is still stepping from one state of symmetry to another. The difficulty in computing a non-computable (52,53) is familiar from the three-body problem (54). Customarily directional transitions are accounted for by a non-Abelian gauge theory (55) but when the Lagrangian is forced to remain invariant, no net evolution will take place but transformations formalize merely the to-and-fro flows of energy. Therefore no continuous group of transformations will account for the natural processes and no Lagrangian will remain invariant during evolution.

During evolution the landscape's curvature will decrease when the spatial potentials $\mu_j = \partial_{Nj}U_{jk}$ and $\mu_k = \partial_{Nk}U_{jk}$ adjust to accommodate or discard the vector potential that couples the system to its surroundings via the *jk*-transformations. The diminishing curvature of a differentiable manifold, i.e., the force can be represented by a vector field gradient. The non-vanishing Lie's derivative (56) means that the change **v** = $d_t$**x** in the coordinate and the change **F** = $d_t$**p** in the momentum, are not collinear due to the net energy flux $\partial_t Q$ over *dt* to the system from the surroundings or *vice versa*. Operators in $[\hat{p}, \hat{x}] = -i\hbar$ do not commute by the minimum amount of action in a change of state which is also inherent in any detection. Consequently the open, non-parallel actions cannot be ranked since these non-Euclidean varieties are without equivalence relations.

The evolving energy landscape is represented by

$$d_t 2K_{\mu\nu} = F^{\mu\nu} v_{\mu\nu} = \begin{pmatrix} 0 & -v_x \partial_x U & -v_y \partial_y U & -v_z \partial_z U \\ \partial_t Q_x & 0 & v_y R_z & -v_z R_y \\ \partial_t Q_y & -v_x R_z & 0 & v_z R_x \\ \partial_t Q_z & v_x R_y & -v_y R_x & 0 \end{pmatrix} \quad (3.6)$$

where the 4-vector velocity $v_\mu = (-c, v_x, v_y, v_z)$. The flow tensor contracts to the 0-form $d_t 2K = \Sigma d_t 2K_{\mu\nu} = -\mathbf{v}\cdot\nabla U + \partial_t \mathbf{Q} + \mathbf{v} \times \mathbf{R}$ where the change in the kinetic energy balances the changes in the scalar potential due to matter flows as well as the changes in the vector potential due to radiation fluxes. When the system communicates with its surroundings exclusively via radiation, Eq. 3.6 is familiar from Poynting's theorem (57) where the radiation **Q** at the speed of light *c* dissipates orthogonal to the source moving down along –∇*U* at velocity **v** (Fig. 2) (14). Conversely, when the system is stationary $\partial_t \mathbf{Q} + \mathbf{v} \times \mathbf{R} = 0$, its stable orbits are governed by $\partial_t 2K + \mathbf{v} \cdot \nabla U = 0$ which is integrable to the virial theorem $2K + U = 0$ or differentiable to the equation for standing waves.

The energy flow from the density $\phi_j$, which defines its spatial locus $x_j$, to $\phi_k$, which in turn defines $x_k$, is identified as the flow of time (14). Thus the notion of time presupposes the notion of space (58). The motion down along the spatial gradient is irreversible when emitted quanta escape forever and reversible when quanta are reabsorbed. Emission will change the coordinate of the source relative to the sink, and absorption will change the coordinate of sink relative to the source (or vice versa) in accordance with the general principle of relativity. Conversely, when the system is stationary, so are its surroundings. It is familiar from unitary transformations that the steady phase velocity $d_t\varphi = \omega$ does not suffice to distinguish the systemic motions from the surrounding motions in accordance with the special principle of relativity.

## 4. The preon action

Evolution from one stationary-state symmetry to another implies that actions are quantized because the stationary states' conserved currents are on closed orbits, and all bound curves are modular. The divisible circular group means a periodic orbit, and a rational winding number is equivalent to a mode-locked motion. According to the Taniyama-Shimura conjecture, there is for every elliptic curve a modular form of Dirichlet L-series $\Sigma \chi_n n^{-s}$ (59). Its analytical continuation in the complex plane $s = \sigma + i\tau$ is, for characters $\chi_n = 1$, Riemann zeta function $\zeta(s) = \Pi_p(1 - p^{-s})^{-1}$.

Each term in the product is invertible to a formal power series from $r = 1$ to $p − 1$ over the number field. When the product's any-one term $1 − z_p^{-qp} = 0$, where $z_p^{-q}$ is the primitive $p^{th}$ root of unity of index $−q$, then $\zeta(s) = 0$. In other words, the characteristic equation $\det(\mathbf{I}\lambda − \mathbf{U}) = 0$ of $\zeta(s)$ can be solved provided that there is a unitary group generator $\mathbf{U}$ with all eigenvalues $\lambda$ of absolute value 1. The unitary condition $_{r=1}\Sigma^p U_{qr}^* U_{qr} = {_{r=1}}\Sigma^p p^{2\sigma-1} = 1$ complies with the Euclidean norm and defines the roots to be at $\zeta(½ + i\tau) = 0$ in agreement with Riemann hypothesis (29,60). Physically speaking, the state is stable when the corresponding action is modular, i.e., quantized over the closed least-action path.

The curve with zeros (nodes) is familiar from the Bohr model where angular momentum $L = px = 2K\tau = n\hbar$ equals the elementary action $\hbar$ in $n$ multiples. The kinetic energy $2K$ within the period $\tau$ is distributed on the closed orbit in modules enumerated by the principal quantum number $n$. Thus a step of evolution from one stationary state to another is mathematically speaking a step in the modulus of the cyclic group. When the sum of points on the Noetherian ring changes, so also its divisor will change. This is explicit in Eq. 3.1 and implicit in Eq. 3.2 where the dissipation-driven evolutionary step of $N_j$ will in fact cause the action to step in momentum $p$ and in length $x = vt$ from one closed orbit to another.

The constant of action $\hbar$, as the absolutely least action, can be considered both the fundamental element of space and the fundamental element of time (Fig. 3). The elementary fermion embodies the energy density of the bound geodesic given by the geometric product $\mathbf{L} = \mathbf{px}$ (61) which has a specific handedness, usually referred to as spin ±½. This oriented element of space is equal to the absolutely least angular momentum $\mathbf{L}$ associated with the kinetic energy within the orbital period $2K\tau = \hbar$. During the transformation from one chiral loop to the opposite handedness, the elementary fermion opens up and becomes momentarily the elementary boson that closes anew but in the reverse sense of circulation. The most elementary boson carries the energy along the open directed geodesic $\mathbf{pv}t$ which has a specific handedness, usually referred to as polarization ±1. This oriented element of time is equal to the absolutely least action $2Kt = \hbar$ that contains the energy carried within the wave's period. It takes two $\hbar$-actions to reverse the polarization from +1 to −1. The first will interfere destructively (head-on) with the original handedness and the other will create the mirror hand. In the following we will refer to the fundamental oriented element of time as the photon $\gamma$ and its opposite sense of polarization as the antiphoton $\gamma^*$. Accordingly, we will refer to the fundamental oriented element of space as the neutrino $\nu$ and its opposite sense of circulation as the antineutrino $\nu^*$. According to the physical portrayal of nature, the constant of action $\hbar$ is the most elementary action which is abbreviated here as the preon (62).

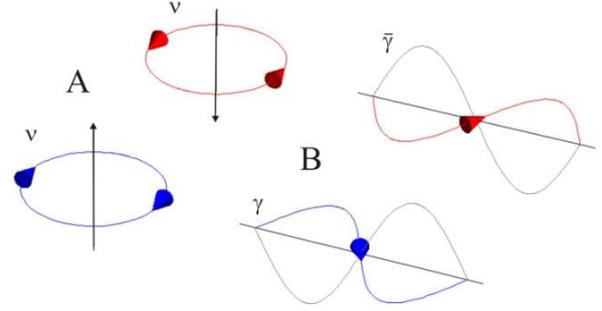

Figure 3. (A) The basic element of space is the most elementary fermion, the neutrino. The confined circulation exists in two chiral forms $\nu$ and $\nu^*$ corresponding to the opposite senses of circulation. The vertical bars denote the respective scalar potentials $-U = 2K$ and $U = -2K$. (B) The basic element of time is the most elementary boson, the photon. The open flow of energy exists in two forms $\gamma$ and $\gamma^*$ of polarization corresponding to the opposite (color-coded) phases $Q = 2K$ and $-Q = -2K$. (The figures were drawn with Mathematica 7 that was appended with CurvesGraphics6 written by G. Gorni.)

## 5. Multiple actions

All spatial and temporal entities, diverse fermions and bosons (63) can be pictured as being ultimately composed of the preon actions. Since the action is a directed path, each fermion and each boson is distinguishable from its own antiparticle which is the reversed action. An electron $e^-$ is figured as a least action path where preons coil to a closed torus having the electron neutrino $\nu_e$ chirality (Fig. 4). Due to the helical pitch the self-generated electromagnetic field deviates from the axis of circular polarization and drives the curved path to the toroidal closure. Electron's steady-state characteristics are obtained from $d_t L = 0$. This resolves to a constant $2K = \int \rho \mathbf{v} \cdot \mathbf{E} dt = \int \rho \mathbf{E} \cdot d\mathbf{x} = e^2/4\pi\varepsilon x$ where the density $\rho$ distributes on the torus' path length $x$ so that the conserved quantity, known as the elementary charge $e = \rho x$, sums from the Noetherian current in the field $\mathbf{E}$ according to Gauss's law. The invariant fine structure identifies by integration to the normalized constant $L/\hbar = \int 2Kdt/\hbar = e^2(\mu/\varepsilon)^{½}/2h = \alpha$ where the squared impedance $Z^2 = \mu/\varepsilon = (c\varepsilon)^{-2}$, in turn, characterizes the stationary-state density that satisfies the Lorenz gauge $\partial_t\phi + c^2 \nabla \cdot \mathbf{A} = 0$ (64). Moreover, the magnetic moment $\mu_e = \int \mathbf{r} \times \rho \mathbf{v} dx = e\mathbf{r} \times \mathbf{p}/m_e = e\mathbf{L}/m_e$ results from the finite torus. Its anomaly $\alpha/2\pi$ (65), i.e., the excess of $\mu_e$ over $\mu_B = e\hbar/m_e$ follows from the helical pitch. The coiling contributes to $\mu_e = eI\omega/m_e = ex^2/t$ via the torus path length $x$ beyond the plain multiples of $\hbar$ (as if the path were without pitch) where inertia $I = m_e x^2$ and $\mathbf{L} = I\omega$, (Fig. 4).

As well, the positron $e^+$ is the torus with antineutrino $\nu_e^*$ handedness. In energy-sparse surroundings $e^+$ and $e^-$ interfere destructively apart from the residual modulation due to their opposite helical pitches. Thus, the annihilation bursts out anti-parallel modulations $\gamma$ and $\gamma^*$ each equivalent to the characteristic mass $m_e = 511$ keV/$c^2$ of the elementary charge (Fig. 4). In topological terms, the low mass means that the winding number of the elementary charge about the torroid's center is low. When dissipation is normalized by the quantum of action, the characteristic frequency $\omega = 2m_e c^2/\hbar$, referred to as the Zitterbewegung (66), is obtained.

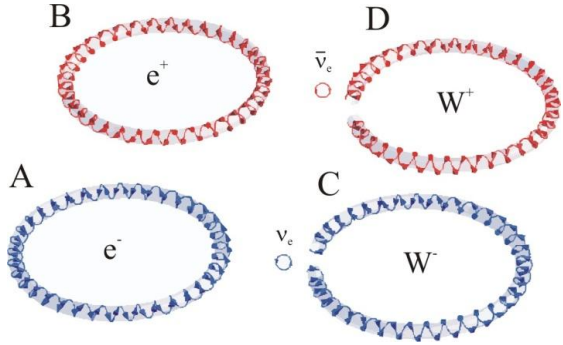

Figure 4. (A) Electron $e^-$ is a closed torus wound of multiple preon actions having neutrino sense of chirality. Due to the helical pitch the lag-phase $\varphi = 2\pi$ is accrued around the torus as seen from the roll of the arrow heads. (The pitch is exaggerated for clarity). (B) Conversely, positron $e^+$ is a closed torus of multiple preons of antineutrino chirality. (C) At high surrounding energy density, $e^-$ converts by breaking its closed chiral symmetry to the open $W^-$ boson and electron neutrino $\nu_e$. (D) Conversely, $e^+$ converts to $W^+$ boson and electron antineutrino $\nu_e^*$. The opposite actions annihilate to streams of $\gamma$ and $\gamma^*$ that results from the lag-phase modulation accrued along the opposite helical pitches.

The $W^-$ boson is an open-ended helix of $\nu_e$-chirality. The helical world-line in energy-sparse surroundings is not the least action path and decays as $W^- \to e^- + \nu_e^*$. The pitch-accumulate lag-phase $\varphi = 2\pi$ is absorbed at the torus closure by the antineutrino. Since the torus itself is a loop, the elementary charge is conserved. Likewise, the $W^+$ boson is an open-ended helix of $\nu_e^*$-chirality that processes as $W^+ \to e^+ + \nu_e$. The neutral $Z^0$ boson is also an open-ended path where a $\gamma$-linker joins two helices one having $\nu_e$ and the other $\nu_e^*$-chirality. Energy-sparse surroundings drive $Z^0 \to e^+ + e^-$. The weak bosons display extraordinary high masses in comparison to their closed form fermion-antifermion counterparts, because these open paths have comparatively little topological self-screening via intrinsic phase cancellation, i.e., their winding numbers are high.

In an atomic nucleus, proton $p^+$ is portrayed as a least action path where two $^2/_3$-helices of $\nu_e^*$-chirality, known as up-quarks u, and one $^1/_3$-helix of $\nu_e$-chirality, i.e., down-quark d, join via three $\gamma$-linkers, known as gluons g. Along each u the helical path accumulates $\varphi = {}^{4\pi}/_3$ and likewise d accrues $\varphi = {}^{2\pi}/_3$ so that a gluon, as an open preon, links any two quarks at the angles that the faces of an equilateral tetrahedron make with each other (Fig. 5). The front-end $\odot$ of one u links via g to the back-end $\otimes$ of the other u, and then further $\odot$-u links via g to $\otimes$-d and finally $\odot$-d links via g to $\otimes$-u to close the path. In the tripod constellation each quark as an element of the closed path is distinguishable from any another which is the essence of quantum chromodynamics based on the non-Abelian SU(3) group.

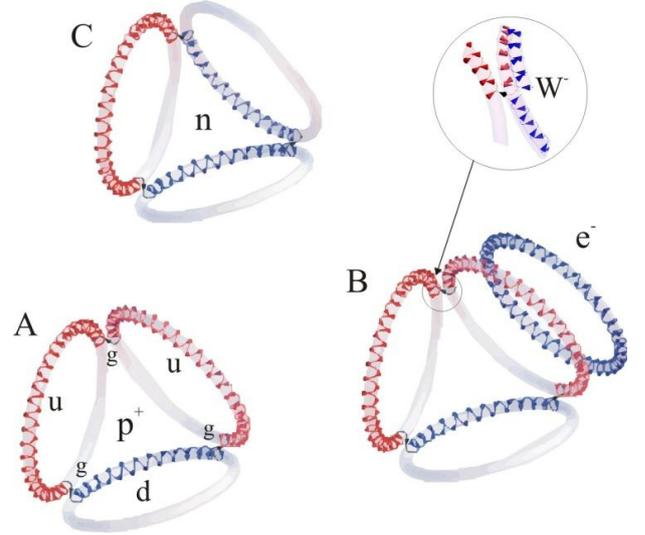

Figure 5. (A) Proton $p^+$ is a closed circulation where each up-quark u (red) is $^2/_3$ of torus of $\nu_e^*$-chirality and the down-quark d (blue) is $^1/_3$ of torus of $\nu_e$-chirality. The lag-phase $\varphi = {}^{4\pi}/_3$ accrued along each u and $\varphi = {}^{2\pi}/_3$ along d due to the helical pitch, define the relative angles of quarks, i.e., symmetry of the closure linked by gluons g (black arrows). (B) Electron capture $p^+ + e^- + \nu_e^* \to n$ is intermediated by $W^-$ which initiates a partial annihilation at an exposed end of u (blow-up) and will yield d. (C) The resulting neutron n is a three-gluon-linked closed action of three quarks udd.

Proton $p^+$ transforms to neutron n via electron capture where $e^-$ breaks, when attracted to u, to $W^-$ which make a snug fit at u so that a partial annihilation will commence and yield d (Fig. 5). Likewise, when n is free, i.e., in energy-sparse surroundings, the reverse process begins when $W^+$ is attracted to make a snug fit at d and the partial $W^+$d-annihilation will yield u. The small photon efflux powers the slow natural process having a long life-time. The provided "wire frame models" are illustrative but perhaps puzzling since $p^+$ by the perimeter is longer than n, yet the mass of the neutron is slightly bigger than that of a proton. However as clarified earlier above, the mass is a measure for the net dissipation. The topological self-screening of u in $p^+$ and d in n are nearly the same. The small difference is accrued from the incomplete cancellation of opposite phases of pitch. The significant difference in proton and neutron magnetic moments, in turn, is understood to stem from the substantial differences in the oriented areas that are closed by the respective currents.

The wire frames allow an easy imagination of various natural process such as a pion decay $\pi^+$: $ud^* \rightarrow W^+ \rightarrow e^+ + v_e^*$, where the $ud^*$-ring opens so that $d^*$, as the oppositely wound d, will resettle via the high-mass intermediate $W^+$ integrally to the low-mass $e^+$-torus. On the other hand the proton decay as a putative process $p^+ \rightarrow e^+ + \pi_0$ (67) is figured in terms of actions so that the X-boson (balanced by $e^+ + d^*$) would be attracted to $p^+$ to make a snug fit along u-⊗-g-⊙-u. However, such an annihilation process is unlikely to take place in low-density surroundings (Eq. 2.1) because it would yield disjoint d and $d^*$ which are high-energy by-products. The process does not yield the anticipated pion $\pi_0$ which would indeed have only moderate mass since in the meson, the circulations of linked quark and antiquark pair cancel apart from the lag-phase accrued along their opposite pitches. Thus, to violate color confinement would require extraordinarily high-energy surroundings to provide the asymptotic freedom (68) for quark-gluon plasma, or to violate baryon number conservation would require a blazing transformation (69,70).

Heavier fermions are excited strings of the aforementioned ground-state actions. An excitation lifts the elementary closed symmetry SU(2) so that an action at a higher harmonic (bending) mode has more to dissipate and thus manifests itself as a heavier particle that tends to decay back to ground state. Altogether the collective modulations of the preon multiples are known in the Standard Model as flavors of leptons, quarks and mesons. The probability amplitudes of the flavor-conserving oscillations are given by the diagonal elements and the flavor-exchanging, familiar from kaon and neutrino oscillations, by the off-diagonal elements of a unitary matrix, specifically CKM for pairs of quarks and by PMNS for leptons. Flavor combinations yield diversity of baryons and mesons. Moreover, the physical portrayal of fermions as closed actions, clarifies that composite actions of spin values above ½ mean networked circulations that give rise, e.g., to quadrupolar moments.

## 6. Fundamental forces

No fundamental interaction is basically different from any other, when everything is viewed as being ultimately composed of multiple preon actions. Energy dispersal only manifests itself differently at different densities. High-density actions are highly curved. A strong force is required to open these closed actions, i.e., to turn a fermion to a boson. In this sense for every fermion there is a corresponding boson which is the familiar notion from supersymmetry. At high nuclear densities quarks integrate seamlessly via gluons to a closed path. Internal reflections at density boundaries, familiar from optics, presumably confined as well the very high density of the nascent Universe. Likewise, a weak force is sufficient to open lepton curvatures to weak bosons and photons.

The Coulomb inter-action, in turn, manifests itself between actions that generate charge. At the thermodynamic steady state the bosons would form standing waves between the oppositely charged fermions. For example, at intermediate densities a bound action known as the hydrogen atom results from the γ-exchange between $p^+$ and $e^-$ (Fig. 6). When the surrounding density increases, the modular paths along the field lines will extend up to an excited state. Conversely, when the surrounding density settles down again, the sparse surroundings will accept the discarded full-wavelength modules of density as the system returns to the ground state. In contrast, like charges, which generate fields of like polarization, cannot support standing waves between them. Therefore repulsion remains the way to dilute the density between them toward that of the sparse surroundings.

Moreover, it is important to realize that the density due to the four-potential (Eq. 3.4) exists beyond net neutral bodies even when the resultant electromagnetic field of anti-phase waves, i.e., the force experienced by charges, vanishes. An effect of the non-vanishing density is familiar from the Aharanov-Bohm experiment (14).

The gravitational inter-action, just as with other interactions, results from density differences. Two isotropically net neutral bodies are attracted when their sparse surroundings, characterized by $\mu_o\varepsilon_o = c^{-2}$ and $\mu_o/\varepsilon_o = Z^2$, accept actions that are released in transitions from one to another standing-wave orbit (Fig. 7). The gravitational force is the difference between the density between two celestial bodies and their surrounding density just as it is the difference between the density in a microscopy cavity and its surrounding density. A cavity, tiny or enormous, cannot accommodate all those standing modes that are propagating in the universal surroundings (71). The fundamental modulus of a standing gravitational wave is a pair of anti-phase bosons. The expression of a standing wave as a sum of oppositely propagating waves subject to boundary conditions (72) is familiar from the derivation of Planck's law (73). Since a modulus of density wave between net neutral bodies is without a longitudinal vector character, a traceless tensor (cvf. electromagnetic wave) denotes transverse modes of the scalar, known as the graviton G. It takes four $\hbar$-actions to reverse the polarization of this Goldstone particle from +2 to –2. Two preons will interfere destructively with the pair with original handedness and the other two will create the mirror-handed pair.

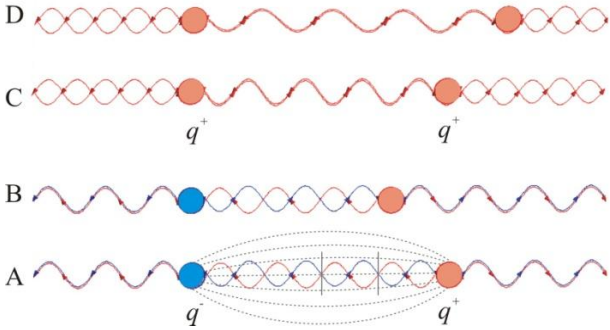

Figure 6. (A) Electromagnetic interaction arises from the density difference between charged fermions and their surroundings. (B) Sparse surroundings cause attraction by accepting full wavelength modules $|\infty|$ of photons from the bound actions that pair opposite charges $q^-$ and $q^+$. For clarity only the shortest action along the electromagnetic field lines (dashed) is decorated with density modules. Beyond the bound pair of charges energy density propagates at the speed $c^2 = 1/\mu_o\varepsilon_o$ and gives rise to the impedance $\mu_o/\varepsilon_o = Z^2$ also long the dipole axis where the anti-phase waves do not couple to the antenna. (C) Like charges generate alike-polarized, open actions that cannot pair to a bound action. (D) Therefore in sparse surroundings repulsion remains the mechanism to dilute the density between $q^+$ and $q^+$.

The familiar form of the conserved potential $U$ for two net neutral bodies displaced by $\mathbf{r}_{12}$ is obtained, as before, from the steady-state condition which is equivalent to the virial theorem $2K + U = 0$ as follows

$$d_t\mathbf{L} = d_t\mathbf{p} \times \mathbf{r} = 2K = \omega^2 m_1 r_1^2 = \omega^2 \frac{m_1 m_2 r_{12}^2}{M} \qquad (6.1)$$
$$= \frac{R^3}{Mt^2}\frac{m_1 m_2}{r_{12}} = \frac{1}{4\pi\rho t^2}\frac{m_1 m_2}{r_{12}} = G\frac{m_1 m_2}{r_{12}} = -U$$

where the angular momentum $L = I\omega$ depends on inertia $I = \Sigma m_i r_i r_i = \Sigma m_i m_j r_{ij}^2/\Sigma m_i = \Sigma m_i m_j r_{ij}^2/M$ and period $\tau^2 = 1/\omega^2 = t^2 r^3/R^3$. The inertia, when normalized by the total mass $M$ of the Universe, can be seen as an expression of Mach principle $4\pi G\rho t^2 = 1$ where the mass density $\rho$ is within the radius $R = ct$ (74), so that any given density is coupled to every other density. The numerous (decay) paths for the energy dispersal span the affine manifold where the density flows are leveling off any density difference in the universal buoyancy in the least time. Thus the cosmological principle is a consequence of a natural selection for the maximal dispersal (29).

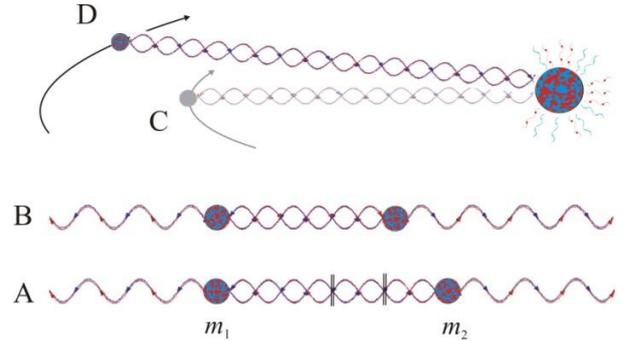

Figure 7. (A) Gravitational interaction arises from the energy density difference between net neutral bodies and their surroundings. (B) Sparse surroundings cause attraction between the bodies by accepting doubly paired density modules $||\infty||$ of gravitons from the bound actions that pair the bodies along gravitational lines. (C) Irradiative combustion of high-density actions is a powerful mechanism to diminish the density difference between a body and its surroundings. (D) The burning star gives away density, hence the conservation of energy (Eq. 3.2) requires for a planet to advance its perimeter from one modular orbit to another.

According to the 2$^{nd}$ law of thermodynamics a system may emit quanta when its surroundings are lower in energy density. The ultimate sink is the universal surrounding free space. Today, at $t = 13.7$ billion years, the very low average mass density $\rho \approx 0.8 \cdot 10^{-27}$ kg/m$^3$ (75) corresponds to a tiny curvature over a titanic radius $R = ct$ (76,77,78) and sets the gravitational constant $G = 1/4\pi\rho t^2$ to its current value. The minute but non-negligible acceleration $a_t = GM/R^2 = 1/\varepsilon_o\mu_o R = c^2/R = c/t = cH$ drives the on-going expansion at the rate $H$ (79,80) that is changing as $d_t H = -H^2 = -4\pi\rho G$. On the basis of $\rho$ or equivalently of $H$ the dilution factor $n = L/\hbar$ is on the order of $10^{120}$. This value quantifies the well-known discrepancy to the invariant Planck scale as if no expansion had taken place. At any given time the total mass of the Universe in its dissipative equivalent $Mc^2 = Ma_t R = R\partial_r U = -U$ matches the scalar potential. This innate equivalence is known also as the zero-energy principle (6). The balance is also reflected in the quantization of the electromagnetic spectrum. The cosmic microwave background radiation profiles according to the Planck's law a discrete quasi-stationary entropy partition of fermions (Fig. 1). The unfolding Universe is stepping from one mode to lower and lower harmonics (81) along its least action path toward the perfect, i.e., torsion-free flatness $d_t L = \tau = 0$. The acceleration, which is proportional to the winding number $\theta \propto R^{-2}$ of those folded actions, will be gradually limiting toward zero. When all densities have vanished, no differences, i.e., forces will exist either.

The Universe has today evolved to span numerous levels of hierarchy but there is no fundamental distinction between the diverse forms of energy, e.g., light, quarks, atoms, molecules, beings, planets, solar systems and galaxies, since they all are ultimately composed of preons. The vast variation in energy densities is reflected in the relative strengths of interactions that extend over some 38 orders of magnitude but there is no profound distinction between strong, weak, electromagnetic and gravitational interactions. All forces as density differences between the preons in diverse fermionic varieties are communicated by the preons in various bosonic forms. At any level of nature's hierarchy, e.g., from neutrino via electron to an atom, a closed least action affords at most three dimensions of space which are usually interpreted as the degrees of freedom. One axis is orthogonal to the plane of the other two where the least action curve exists and closes to a cycle in accordance with the Poincaré-Bendixson theorem (82). Indeed it makes sense to see space at higher and higher magnifications as being folded more and more [as highly curved actions], but not to imagine the degrees of freedom as an abstract concept without physical correspondence. In turn, an open action associates with the degree of freedom that span the dimension interpreted as time. When the system evolves from one spatial state to another, open actions carry the flow of energy that is experienced as the flow of time. It has unique direction due to the natural process that diminishes density differences. At a thermodynamic steady state in- and outgoing bosons are equally abundant. When the two populations of polarizations form a standing wave, motions are fully reversible. There is no time-dependence but time as a bidirectional parameter has two degrees of freedom. Indeed it makes sense to see time when broken down to finer and finer steps, as a quantized energy transduction process but not to regard irreversibility and reversibility as abstract notions without physical correspondence.

In addition to those examples described above the principle of least action serves to portray many other intriguing phenomena. For example, a laser is a (nearly) isolated system where an optical cavity closes one or more standing modes. Likewise, a superconductor is a system that is isolated from its surroundings by an energy gap and where standing vibrations of lattice pair electrons of opposite spins to coherent states. In turn, a condensate is so sparse with bosons that fermions fail to keep distinguishing from each other by populating excited states. Instead, they can only afford to pair by exchanging bosons, i.e., $\Delta Q_{jj} = 0$ producing indistinguishable isergonic actions of modulus $N_j$.

## 7. The mass gap

Evolution is basically energy dispersal by spontaneous symmetry breaking. Although it is obvious, it is important to note that any transformation from a closed orbit to an open curve is discontinuous. In other words, the action is either closed or open but not anything in between. No change of symmetry is indivisible and no energy spectrum is continuous. Eventually, when evolution arrives at the state of next most elementary symmetry, the closed preon opens up and drains altogether as an open preon (Fig. 8). The winding number of the most elementary non-commutative Noetherian ring is ±1 whereas for an open wave it is zero. Thus the least mass $m_\nu > 0$ corresponding to the Noetherian charge of the closed preon is associated with the SU(2) symmetry whereas the open preon associated with U(1) is without mass $m_\gamma = 0$. The most elementary symmetry cannot be broken any further. For that reason there is no primitive root of unity, i.e., a standing-wave solution corresponding to a mass. It follows that there is a mass gap which is a finite difference in energy between the lowest bound state and the free, open state.

It is common to denote by a vacuum vector $\Omega$ the eigenstate of Hamiltonian $H$ with zero energy $H\Omega = 0$, but this notion is abstract without physical correspondence. When there is no energy density, there is no state either. The surrounding open free space of the Universe is not empty but the energy density known as the vacuum amplitude, in some theories referred to as the Higgs field, is in balance with its fermion systems (Fig. 1). This is apparent from the cosmic microwave spectrum that matches the black-body spectrum. The conservation of energy requires that at any given step during the universal energy dispersal process, the change in kinetic energy density, deemed as continuous in $d_t(\rho v^2) = -\mathbf{v}\cdot\nabla u - \varepsilon_o c^2 \nabla\cdot(\mathbf{E}\times\mathbf{B})$, balances the change in the scalar potential density $u$ due to the average density $\rho$ of fermionic matter with the change in the bosonic electromagnetic radiation density. The Poynting vector $\mu_o^{-1}\mathbf{E}\times\mathbf{B}$ (57) embodies the energy of free space which is unique up to a phase and invariant under the Poincaré group in accordance with Maxwell equation $d\mathbf{F} = 0$ in the absence of charges and currents.

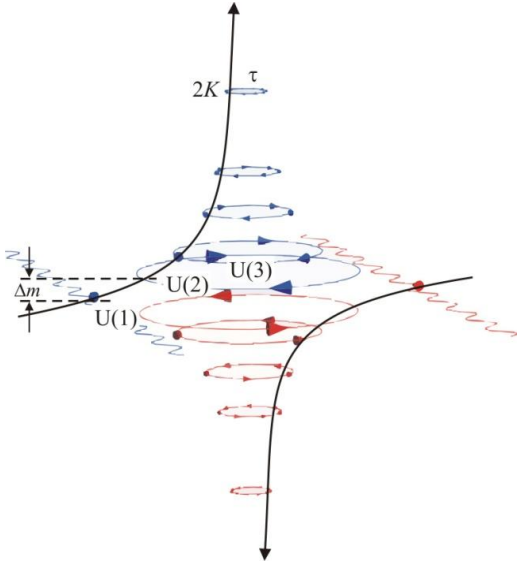

Figure 8. Level diagram depicts an array of actions (colored) that are classified according to unitary groups of symmetry. For each stationary action the kinetic energy $2K$ integrated over the period $\tau$ along a directed path associates via Noether's theorem with a conserved quantity $m$. A mass gap $\Delta m \neq 0$ exists between the closed action of the most elementary fermions, the neutrinos defined by the next lowest symmetry group U(2) and the open action of the propagating photons defined by the most elementary group U(1). The radiated energy $\Delta E = \Delta mc^2$ in the transformation from U(2) down to U(1) amounts from the photons that immerse in the surrounding free space. Its density is non-zero as is apparent from the finite speed of light. Only one of the two chiral forms of U(2) fermions can be detected via emissive transformations to U(1) whereas a detection by absorption would transform the U(2) particle to another one of a higher symmetry group.

The Lie derivative is conveniently continuous in evaluating the change of one vector field ($\partial_t Q/c$) along the flow of another vector field ($\mathbf{v}\cdot\nabla U$). The smooth, differentiable manifold as a Lie group, such as the gauge group U(1)×SU(2)×SU(3) of the Standard Model, is admittedly suitable for mathematical manipulation, but would be appropriate only for a stationary system because it fails to describe the discontinuous event of breaking symmetry. At the symmetry break the Jacobi identity does not hold. The step from one stationary state symmetry to another takes at least one quantum of action from the system to its surroundings. When the quantized action is understood as the basic constituent of nature, no renormalization is required to escape from singularities that trouble theories based on the energy concept (83). Thus, it is not productive to maintain that the Lagrangian would be invariant and that evolution would be a continuous group of transformations. Accordingly, it is inconsistent to insist on the existence of a theory which would both comply with a gauge group via renormalization, and also display a mass gap (84).

**Theorem**: For any-one compact simple gauge group G, there exists no quantum Yang–Mills theory on $R^4$ that has a mass gap.

*Proof*: The Yang–Mills Lagrangian

$$L = \frac{1}{4g^2}\int Tr\, \mathbf{F} \wedge *\mathbf{F} \qquad (7.1)$$

where the two-form $\mathbf{F}$ denotes curvature via a gauge-covariant extension of the exterior derivative from the one-form $\mathbf{A}$ of the G gauge connection and $g^2$ is the determinant of the metric tensor of the spacetime. It follows that G governs the symmetry of the action, which is the integral of $L$. Since the quadratic form that is constructed from $\mathbf{F}$ and its Hodge dual $*\mathbf{F}$ on the Lie algebra of G on $R^4$ is invariant, so also $L$ is invariant and so is its integral action invariant. The invariant action relates via the Noether's theorem to a quantity $m$ whose value is conserved. It then follows that a difference between any two values $m_G$ and $m_H$ of a conserved quantity cannot be compared within any-one theory based only on a single gauge group. Specifically, in a Yang–Mills theory that is based exclusively on one gauge group G describes only a single, invariant state. When the theory is based on U(1), it describes the vacuum state (the free space). When the theory is based on another Lie group, it describes another state. The difference $\Delta m = m_H - m_G \neq 0$ in the conserved quantity between U(1) and the next lowest symmetry group is referred to as the mass gap. However, the gap is not defined, i.e., it does not exist within any given theory based exclusively on a single gauge group.

**Corollary**: For any-two distinct gauge groups G and H, there exists a theory on $R^4$ that has a mass gap.

*Proof*: Each gauge group relates the symmetry of action as the invariant integral of Lagrangian on $R^4$ via the Noether's theorem to a quantity *m* whose value is conserved. Conversely, the two distinct gauge groups relate two distinct symmetries of invariant actions to two different values $m_G$ and $m_H$ of the conserved quantity. Thus there exists a difference $\Delta m = m_H - m_G \neq 0$ in the conserved quantity, i.e., the mass gap, between the vacuum state energy $E_G$ governed by the most elementary G = U(1) symmetry group and the lowest excited state energy $E_H$ governed by the next most elementary symmetry group H. Since the mass is a measure for the amount radiation that can be dissipated to the surrounding vacuum, the vacuum itself has zero mass.

The resolution of the Yang-Mills existence and the mass gap, while trivial in its proof, is revealing. The focus is not on energy but on action which is quantized for each state according to the corresponding gauge symmetry (Fig. 8). Therefore, no matter how small the energy *E* of an excited state will be decreasing over an increasing time *t*, the product $Et = n\hbar$ remains invariant holding at least one (*n* = 1) multiple of the quantized action. Likewise, no matter how far the range **x** of a field will be extending with diminishing momentum **p**, their product **px** remains invariant so that associated observables remain countable. In other words, a theory based on action is self-similar but does not require renormalization since it is not troubled by infinities and singularities related to energy and time as well as to momentum and length. Yet, it may appear for some to be unconvincing to claim, as above, that SU(2) which is by determinant isomorphic to U(1), does have a distinct property, i.e., mass. However, a unitary group U(*n*) is non-Abelian for *n* > 1 whereas U(1) is Abelian. The sense of circulation distinguishes the closed action from a point, and also from its open form because when the chiral, closed action is mapped on the polarized open action, the two open termini are distinct from the integrally closed and connected path. To bridge a topological inequivalence requires a dissipative transformation process. The mass gap is between the photon and the neutrino. Moreover, the Euler-Lagrange integral of a bound trajectory is well-defined, i.e., countable whereas Maupertuis' action as an open path is ambiguous, i.e., uncountable. The ambiguity is reflected in the assignment of a value to the free energy of the vacuum. Instead the surrounding vacuum characteristics $c^2 = 1/\varepsilon_o\mu_o = GM/R$ are defined via the closure according to the Stokes theorem.

## 8. Algebraic and non-algebraic varieties

In the universal energy landscape a system as a variety composed of varieties is in relation to all other varieties via flows of energy. The status of a system

$$P = \prod_j P_j = \prod_j \left( \prod_k \left( N_k e^{-(\Delta G_{jk} - i\Delta Q_{jk})t/n\hbar} \right)^{N_j} \Big/ N_j! \right) \quad (8.1)$$

given in terms of actions is quantized $L = 2Kt = k_BTt = n\hbar$ (cvf. Eq. 2.2). The stationary state ($d_tP = 0$) measure $\ln P = \Sigma N_j$ is a sum over all closed graphs, as is familiar from index theory (82) and from the linked cluster theorem for reversible, i.e., non-propagating processes (85). The formalism relates also to that of the quantum field theory so that an open action due to the vector potential, denoted as -$iQ_jt$ in Eq. 8.1, relates to the creation of a *j*-action by the boson operator $b_j^\dagger$ and $iQ_kt$, in turn, relates to the concurrent destruction of a *k*-action by $b_k$ according to the canonical commutation relations algebra. Conversely, a bound action due to the scalar potential, denoted as -$U_jt$ in Eq. 8.1, associates with the creation of the *j*-action by the fermion operator $f_j^\dagger$ and $U_kt$, in turn, relates to the concurrent destruction of the *k*-action by $f_k$ as formalized by the canonical anticommutation relations algebra.

It is in the objective of physics to predict by way of calculation. However, as has been emphasized all along, natural processes are in general non-computable. Computation is ultimately ranking, and intuitively it is impossible to rank a quantity that is changing during the computation. Since an evolutionary step means a change in the modulus of an action, formalized as a Noetherian ring, a descending sequence eventually attains the preon ring (infima) in some finite surroundings. Conversely, an ascending sequence attains some maximum number of modules (suprema) at a congruence which corresponds to the thermodynamic steady state in some finite surroundings. The ordering ensures that any bounded space is compact. The stationary state's conserved character is also familiar from the fundamental theorem of Riemannian geometry which states that on any Riemannian manifold there is a unique torsion-free metric connection, the Levi-Civita connection. The to-and-fro flows of energy are the affine connections that establish the ordering relations between conserved actions at the net non-dissipative state. These equivalence relations identify the countable subspaces, known as algebraic varieties, from which new spaces can in turn be constructed algebraically. Specifically any two varieties are equivalent if and only if they are on the same orbit. The stationary state varieties form an orbit space. Conversely, the finiteness properties ensure that when the orbit space, an $n$-dimensional oriented and closed manifold $M$, is divided in two, the system's group of $k$ varieties and to the surrounding's group of $n$-$k$ varieties are duals $H^k(M) \simeq H^{n-k}(M)$. The Poincaré duality theorem states that for $M$ the $k^{\text{th}}$ cohomology group of $M$ is isomorphic to the $(n − k)^{\text{th}}$ homology group of $M$, for all integers $k$.

In general, the correspondence between the system and its surroundings is stated by the Hodge conjecture so that de Rham cohomology classes are algebraic, i.e., they are sums of Poincaré duals of the homology classes of subvarieties. The conjecture says in terms of physics that stationary systems can be algebraically constructed from subsystems that are duals of the stationary surroundings' subsystems. Also functional equations reflect the Poincaré duality. Since at the conserved stationary state there are no open actions, all Hodge classes are generated by the Hodge classes of divisors where a divisor on an algebraic curve is a formal sum of its closed points on the basis of the Lefschetz theorem on (1,1)-classes. All closed curves with zeros (nodes) are modular which is the essence of Taniyama-Shimura conjecture that all rational elliptic curves are modular (59). Since an algebraic variety is the set of solutions of a system of polynomial equations (elliptic curves), an algebraic number field has a norm and at least one zero. It can be considered as a geometric object of the affine manifold (86). Since every non-constant single-variable polynomial with complex coefficients has at least one complex root, the field of complex numbers is algebraically closed. Moreover, Hilbert's Nullstellensatz relates ideals of polynomial rings to subsets of affine space.

The stationary energy landscape with its closed, oriented and unitary structure is a Kähler manifold $M$ that can be decomposed to its cohomology with complex coefficients, corresponding to the scalar and vector potentials, so that $H^n(M, \mathbb{C}) = \oplus H^{p,q}(M)$ where $p + q = n$ and $H^{p,q}(M)$ is the subgroup of cohomology classes that are represented by harmonic forms of type $(p, q)$. It is instructive to consider the (twice) differentiable and connected $M$, because an $n$-form $\omega$, i.e. the Lagrangian, can be paired by integration, i.e., via the action $\int \omega = \langle \omega, [M] \rangle$ with the homology class $[M]$, i.e., the fundamental class whose the top relative homology group is infinite cyclic $H^n(M, \mathbb{Z}) \simeq \mathbb{Z}$. The action $\int \omega$ over $M$ depends only on the cohomology class of $\omega$.

The physical insight to the mathematical problem of non-countability is consistent with the fact that the Hodge conjecture holds for sufficiently general and simple Abelian varieties such as for products of elliptic curves. Moreover, the stationary least action is consistent with the combination of two theorems of Lefschetz that prove the Hodge conjecture true when the manifold has dimension at most three which is required for the stability of the free energy minimum varieties.

Although the Hodge conjecture complies with conservation and continuity, it is ambiguous in how the duality came about. It is in essence asking an evolutionary question, which cohomology classes in $H^{k,k}(M)$ form from the complex subvarieties? Of course, it has been found mathematically that there are also non-algebraic varieties. For example, when the variety has complex multiplication by an imaginary quadratic field, then the Hodge class is not generated by products of divisor classes. These troublesome varieties correspond to the open actions. They are without norm, hence non-modular and indivisible. The $n$-form $\omega$ of a non-modular curve over a finite field is non-intergrable because it is without bounds. Although the natural processes terminate at the irreducible open preon, the path is open because one photon after another leaves the system. The manifold keeps contracting, mathematically without a bound, but physically the landscape ceases to exist when the last fermion opens up and leaves forever. Also singularities associated with the squared operators, e.g., the square of an exterior derivate, are troublesome abstractions. At a singular point the algebraic variety is not flat which means it is non-stationary and uncountable.

Thus the proof of the conjecture that Hodge cycles are rational linear combinations of algebraic cycles, hinges on excluding non-integrable classes, i.e., those without divisors. This depends on the definition of manifold.

**Conjecture**: Let $M$ be a projective complex manifold. Then every Hodge class on $M$ is a linear combination with rational coefficients of the cohomology classes of complex subvarieties of $M$.

*Proof*: A divisor on an algebraic curve is a formal sum of its closed points. It follows that if $M$ accomodates any Hodge class that cannot be generated by the Hodge classes of divisors, then the conjecture must be false. The definition of a projective complex manifold as a submanifold of a complex projective space determined by the zeros of a set of homogeneous polynomials, excludes any indivisible polynomial without roots (87). Since transition functions between coordinate charts are by definition holomorphic functions, also singular points are excluded from $M$. The coordinate chart is a map $\phi: U \to V$ from an open set in $M$ to an open set in $R^n$ of the dimension of $M$ is one-to-one. The homeomorphism excludes any open curve, i.e., any polynomial without at least one root. Thus it follows that all Hodge cycles on $M$ are rational linear combinations of algebraic cycles on a projective complex manifold. Since the projective complex manifold has no class without a divisor, the conjecture is true.

Those projective algebraic varieties are also called Hodge cycles which reflects the modular ring structure of fermions. These Noetherians are mathematically manipulated by cup and cap products. Moreover, ideals, most importantly the prime ideals as special subsets of a ring, correspond to the eigenvalues of stationary motional modes.

## 9. Discussion

The value of a natural principle is in comprehending complex as well as simple phenomena in the same basic terms. The principle of least action came first into sight by Fermat, when addressing in particular the least-time propagation of light and later by Maupertuis when rationalizing in general the least-time flows of any form of energy. However, Euler and especially Lagrange found mathematical reasons to narrow the general formulation to computable cases, i.e., to bound and stationary systems. Ever since physics has mostly evolved along reductionist and deterministic tracks nevertheless, there is nothing in the character of physics that would exclude holistic and non-deterministic descriptions of nature.

Physics, just as any other discipline, is enticed by ambiguity in its central concepts, most notably space and time. According to the principle of least action space at any level of natural hierarchy is embodied in closed, stationary actions that evolve with time which, in turn, is embodied in one or multiple quantum of action that are discarded to sparser surrounding actions. Density differences diminish as well when sparse systems are acquiring quanta from their dense surroundings. Curiously though, when there are alternative paths for the flows, the natural processes are non-deterministic because the flows by the very fact of flowing affect the driving density differences. This character of nature may not please our desire to predict but we had better acknowledge it and get acquainted with its basis. Also we need to recognize that an observation is a causal connection via energy transduction process from an object to an observer where indeterminism in outcomes arises from the relative phases of transmitter and receiver motions (14,34).

The worldview provided by the least-time principle recognizes no fundamental distinction between fundamental particles and fundamental forces but regards all structures and processes as having emerged, proliferated and eventually been extinguished in leveling off density differences. According to the 2$^{nd}$ law of thermodynamics the flows of energy naturally select (88) mechanisms inanimate and animate to process flows in the least time (89,90,91,92). Although it is natural to search for symmetry as it relates to a free-energy minimum state, the mathematical beauty as such cannot tell us how it came about and how it will break down. Closed currents as manifestations of symmetry and topological invariance give rise to conserved quantities, most notably to mass which is ultimately valued in open-action equivalents that are accommodated in the thinning of surrounding space – the physical vacuum (93).

The principle of least action accounts for everything by counting preons. This holistic picture of nature is not new (94,95,96) nonetheless the tenet may appear to some as naive, but in return it provides insight to parameters and processes also beyond those analyzed here. Plethora of particles and diaspora of fundamental forces are seen as merely manifestations of energy dispersal not as ultimate causes or consequences. The on-going universal unfolding of strings of actions accelerates with the increasing radius of curvature, just as earlier, when primordial high densities, presumably similar those in present flavors of elementary particles, began to thin out from high-density resonant confinements closed by internal reflections. The chirality consensus in actions that settled in at baryogenesis is seen as to have been an effective means to disperse energy just as the chirality standard of natural amino acids that established itself at prebiotic genesis as an effective means to facilitate energy dispersal (97,98). Moreover, the physical portrayal of nature by diverse actions in multiples of $\hbar$ is tangible in illustrating mathematical problems related to intractability (52,99,100), emergence, change in modularity (29,101,102, 103), symmetry breaking and division in duals which all are troubled by the problem of the uncountable.

Varied writings of Pierre Louis Moreau de Maupertuis reveal that he was outwardly stimulated in applying the principle of least action to decipher puzzles and phenomena in diverse disciplines (104). Although Maupertuis' formulation failed to meet the integrability condition that was insisted upon by his rigorous contemporaries and has been required ever since on diverse occasions, it took time to understand the underlying reason that distinguishes reversible from irreversible and tractability from intractability and that discerns space from time. Undoubtedly the here-presented revision of M[aupertuis']-theory does not cohere with current classiness but calls for revival of intuition in the quest of unity.

**Acknowledgments**. I am grateful to Jani Anttila, Szabolcs Galambosi, Ville Kaila, Samu Kurki, Tuomas Pernu, Tapio Salminen, Stanley Salthe, Jouko Seppänen, Tuomo Suntola, Mika Torkkeli for insightful discussions and valuable corrections.